\documentstyle[aps,prb,preprint]{revtex}
\tightenlines

\begin{document}
\title{Local order and magnetic field effects on the
electronic properties of disordered binary alloys in the Quantum Site
Percolation limit }
\author{D. F. de Mello}
\address{Departamento de F\'{\i}sica, Faculdade de Ci\^encias, \\
Universidade Estadual Paulista (UNESP), \\
C. P. 473, Bauru 17033-360, SP, Brazil}
\author{G. G. Cabrera}
\address{Instituto de F\'{\i}sica ``Gleb Wataghin,''\\
Universidade Estadual de Campinas (UNICAMP), \\
C. P. 6165, Campinas 13083-970, SP, Brazil}
\date{Received \ }
\maketitle

\begin{abstract} 
Electronic properties of disordered binary alloys are studied {\em
via} the calculation of the average Density of States (DOS) in two
and three dimensions. We propose a new approximate scheme that allows
for the inclusion of local order effects in finite geometries and
extrapolates the behavior of infinite systems following {\em
finite-size scaling} ideas. We particularly investigate the limit of
the Quantum Site Percolation regime described by a tight-binding
Hamiltonian. This limit was chosen to probe the role of short range
order (SRO) properties under extreme conditions. The method is
numerically highly efficient and asymptotically exact in important
limits, predicting the correct DOS structure as a function of the SRO
parameters. Magnetic field effects can also be included in our model
to study the interplay of local order and the shifted quantum
interference driven by the field. The average DOS is highly sensitive
to changes in the SRO properties, and striking effects are observed
when a magnetic field is applied near the {\em segregated} regime.
The new effects observed are twofold: there is a reduction of the
band width and the formation of a gap in the middle of the band, both
as a consequence of destructive interference of electronic paths and
the loss of coherence for particular values of the magnetic field.
The above phenomena are periodic in the magnetic flux. For other
limits that imply strong localization, the magnetic field produces
minor changes in the structure of the average DOS.
\end{abstract}


\narrowtext

\section{Introduction}

Considerable efforts have been made in the understanding of
localization properties of electronic states of disordered systems
since the work by Anderson.\cite{Anderson} Being the simplest one to
investigate the above phenomena, the Anderson model is formulated
through a Tight Binding Hamiltonian with diagonal disorder and
hopping $t$ to nearest neighbors. The site energies are randomly
distributed within some range $W$, and $W/t$ is the relevant
parameter to distinguish between the strong and weak localization
regimes. Anderson showed that disorder, in three dimensional systems,
drives a metal-insulator transition, with a mobility edge separating
localized from extended electronic states.

The one parameter scaling theory, introduced by Abrahams {\em et
al.}, \cite {gangof4} predicts that all states are localized for
dimensions smaller than or equal to two. Dimension two then appears
as a marginal case. The one-parameter scaling hypothesis assumes that
the conductance of a finite sample scales in a universal way, and has
been supported by several experiments done in the early
1980s,\cite{LeeRam} and by a large number of numerical
works.\cite{MacKinonn,Kramer,Soukoulis} However, it has been
suggested that the theory breaks down when electron-electron
interactions are included, or when an external magnetic field is
applied.\cite{weak,rev93} Remarkable new experiments on 2D electron
systems,\cite{exp} in samples with much higher electron mobilities
than the previous ones, do show a clear evidence for a
metal-insulator transition at low electron densities, in the absence
of a magnetic field. The effect has been ascribed to the dominant
role of Coulomb interactions, and a new scaling theory has been
proposed for interacting electrons,\cite{miranda} leaving open the
possibility of a 2D metal.

Within the one-electron scheme, it was well known that in 2D, the
effect of back scattering from impurities is enhanced, leading to an
insulating behavior at zero temperature. Electron trajectories
interfere constructively in the backward direction, and one gets
localization for any degree of disorder. Transport properties with an
external magnetic field are useful probes to test the above effects.
Quantum interference of electron trajectories, modified by the
applied field, may change the localization properties of the electron
wave function. One evidence of this fact is the observation of
magnetoresistance oscillations with an applied field in thin
disordered metallic cylinders.\cite{Sharvin,Altshuler} With this
geometry, and orienting the field along the cylinder axis, one can
select the phase of the electron orbits in order to observe the
phenomenon macroscopically.  For two-dimensional systems, like the
ones realized in semiconductor heterostructures, it has been observed
a metal-insulator transition induced by the magnetic
field.\cite{Wang} Samples that, at sufficiently low temperatures and
zero field, exhibits a localized electronic behavior, present
evidence of extended states at strong magnetic fields. Other
evidences for a transition to a metallic regime have been observed in
experimental studies of the magnetoresistance and
magnetoconductance.\cite {Shih,Katsumoto} As a general rule, quantum
interference effects are invoked to explain the delocalization of the
electronic wave function with the field. \cite{rev93} The reentrant
behavior of the magnetoresistance found by some authors at very high
magnetic fields,\cite{Spriet,Jiang} is associated with the shrinkage
of the electronic wave function.\cite{Efros}

One effect marginally taken into account, and that may have important
consequences, even within the one-electron picture, is the inclusion
of Short-Range Order (SRO) correlations in the electronic properties
of disordered systems.\cite{Mello,Sieg} Real disordered systems are
not necessarily random, but may present different degrees of
short-range order.  In a binary alloy, for given concentrations of
the atomic species, one may get configurations ranging from
segregation to complete order as a function of the SRO properties. In
this work we investigate the role of the above effects and its
interplay with the action of an applied magnetic field, in The
Quantum Site Percolation limit.  The latter regime corresponds to the
case where the two site energies, for a binary alloy, are very far
apart in the energy scale, and {\em percolation} of the wave function
is only possible through atoms of the same species.  Thus, this limit
favors localization when short range correlates atoms of different
kind as nearest neighbors. We remark that this case is difficult to
calculate within other approximations of nonlocal
character.\cite{Sieg} We calculate the average electronic Density of
States (DOS) using a previously developed method reported in
Ref.\onlinecite{Mello}. Our approach relies on one dimensional
numerical techniques, and is efficiently applied to long stripes and
bars of finite cross section with the idea that two and three
dimensional lattices may be obtained by joining infinite linear
chains. From grounds based on Statistical Mechanics
results,\cite{cilindro} this geometry, generically called as {\em
cylindric}, is expected to converge more rapidly to the bulk limit
than the simpler block geometry.  Indeed, for the disordered case,
the bulk regime can be extrapolated with a relatively small number of
chains. SRO effects are included following the Kikuchi
method,\cite{Kikuchi} where correlations are systematically handled.
The last step in the calculation involves a configurational average,
where the probability of a given configuration is obtained in terms
of the short and long range order parameters.  The electronic DOS
calculated through the above procedure, in spite of being an average,
is rich in structures, and as we will show below, is highly
susceptible to changes in local properties.

Magnetic field effects are considered with the introduction of a
phase in the electronic hopping term\cite{rev93} {\em via} the so
called Peierls substitution.\cite{peierls} This phase is related to
the magnetic flux across an elementary cell, as given by the well
known Aharonov-Bohm effect.\cite{Sakurai} The lattice model approach
adopted in this paper is known to accurately describe the lowest
Landau level in the situation where the disorder broadening is such
that no overlap of magnetic sub-bands occurs. We are also neglecting
electron-electron interactions and electron spin (spin-split Landau
levels), and all the effects found in the present contribution are
exclusively due to disorder in a one-particle lattice model. We found
that magnetic field effects are stronger near the segregation limit,
where one expects weak localization properties and a higher degree of
coherence in the electronic wave function. An oscillatory behavior of
the electronic DOS is observed as a function of the field, with
surprising effects on the structure of the DOS.

Our paper is organized as follows: In the next Section we present the
method to calculate the DOS. As illustrative examples, we show
results for one-component non-disordered samples in two and three
dimensions. The two-dimensional disordered case including SRO effects
is also shown for completeness. In Section III we extend our
calculation to three-dimensional disordered systems in the presence
of an applied magnetic field. Finally, in Section IV we present and
discuss numerical results for several examples of the latter.

\section{A new approach to calculate the DOS}

The computational scheme that will be employed here, was developed
previously by the authors to study the electronic DOS in
two-dimensional disordered binary alloys.\cite{Mello} For
completeness, we discuss in this Section some details of the method
and present the extension to three dimensions in the next Section.

A tight-binding Hamiltonian for a single $s$-atomic-like orbital, in
a hypercubic lattice structure in $d$ dimensions, yields the
following dispersion relation in $k$-space:
\begin{equation}
E\left( {\bf k}\right) =E_0+\sum_{i=1}^d\ 2V\ \cos k_ia\ ,
\label{tbenerg}
\end{equation}
where $E_0$ is the atomic energy, $V$ is the hopping (kinetic) term
for nearest neighbors in a pure system, and $a$ is the lattice
parameter. The DOS for $d$-dimensional systems can be obtained
recursively from the DOS in $(d-1)$ dimensions, in the form of a
convolution with the DOS of a linear chain
\begin{equation}
{\cal D}^{(d)}(E)=\int dE^{^{\prime }}{\cal D}^{(d-1)}(E^{^{\prime
}})\ {\cal D}^{(1)}(E-E^{^{\prime }}).  \label{DOS}
\end{equation}
This result is shown analytically in the Appendix for two dimensions.
For a finite system with the geometry of a {\em hyper-cylinder}, like
a strip with finite width, the DOS is obtained as a discrete
convolution of the DOS of the linear chain with the distribution of
eigenvalues for the finite transverse section of the cylinder
\begin{equation}
{\cal D}_{Cylinder}^{(d)}(E)=\sum_i\ {\cal D}_{Section}^{(d-1)}(E_i)\
{\cal D }^{(1)}(E-E_i)  \label{findos}
\end{equation}
In finite systems, strong oscillations are present in the DOS,
specially near the points where van Hove singularities are developed
in the infinite limit (in analogy with the so called Gibbs phenomenon
for Fourier series).  The above oscillations are typical size
effects, with the amplitude being damped when the size of the finite
section is increased.\cite{mello-lagos} In Fig.\ref{fig1} we display
the DOS for the square lattice for wide but finite width stripes,
along with the infinite two dimensional case (uppermost figure). The
corresponding three dimensional case (cubic lattice) is depicted in
Fig.\ref{fig2}. We compare finite bar geometries with the infinite
three dimensional limit. The energy scale is given in units of $V$,
the hopping integral. The large amplitude of oscillations are due to
coherence effects in pure samples, and as we will show later, size
effects decay more rapidly in disordered systems. In addition, as
remarked in the Introduction, the {\em cylindric} geometry, with an
infinite dimension along the axis, is expected to converge
considerably faster to the bulk limit than the block geometry. This
latter fact allows the use of relatively small samples to extrapolate
the DOS of disordered alloys. One comment is in order here. As shown
in expression (\ref{findos}), the DOS for a finite system is obtained
as a weighted superposition of one-dimensional densities. The
interchain coupling displaces the center of gravity of the
one-dimensional DOS to the {\em renormalized} value $E_i$, which
takes into account the true coordination number and the degrees of
freedom of the transverse section of the system. The weight in
(\ref{findos}) for each chain is exactly the discrete density of the
eigenvalues $\{E_i\}$.

The above exact analytic results are extended heuristically to treat
disordered systems. In this Section we show the procedure for two
dimensions, \cite{Mello} where the transverse part of the system is a
finite chain.  Local order effects are handled with a single
variable, the so called Cowley's parameter $\sigma$ \cite{Ziman}:
\begin{equation}
{\sigma}=1-\frac{p_{AB}}{c_Ac_B},  \label{sigma}
\end{equation}
with $p_{AB}$ being the probability of having an $AB$ pair, and $c_A$
and $c_B $ the concentrations for species $A$ and $B$ respectively.
With the above definition, $\sigma$ ranges from $-1$ (limit of
complete local order) to $1$ (limit of segregation), passing through
intermediate cases with variable local order that include the random
limit for $\sigma=0$. For Markovian chains, SRO effects were embodied
in a numerically exact algorithm based on the so called Negative
Factor Counting (NFC) method\cite {Dean} to calculate the DOS for
such systems. Our scheme thus relies on exact calculations for ${\cal
D} ^{(1)}(E;\sigma)$, the DOS of disordered infinite chains with
local order described by $\sigma$. To get the analogous of expression
(\ref{findos}) for a disordered two-dimensional stripe of width $M$,
we generate all the configurations of a finite chain of $M$ sites,
and get the corresponding eigenvalues and their multiplicities. For a
given configuration, the DOS is obtained as
\begin{equation}
{\cal D}_{Strip}^{(2)}(E;M,c)=\sum_{i=1}^Md_{Section}^{(1)}(E_i;M,c)\
{\cal D }^{(1)}(E-E_i;\sigma),  \label{strip}
\end{equation}
where the ${\cal D}^{(1)}(E-E_i;\sigma)$ is an infinite chain DOS
with the center of gravity displaced to $E_i$, the eigenvalues of the
finite transverse chain for the given configuration. Each term of the
sum (\ref{strip}) is weighted by $d_{Section}^{(1)}(E_i;M,c)$, the
discrete distribution of the eigenvalues $\{E_i\}$. The final DOS is
an average over all the possible configurations:
\begin{equation}
{\cal D}_{strip}^{(2)}(E)=\sum_cP_c\ {\cal D}_{Strip}^{(2)}(E;M,c)\ ,
\end{equation}
with $P_c$ equal to the probability of a given configuration $c$. The
latter is obtained, for Markov chains,\cite{markov} as a weighted
product of pair probabilities and depends only on $\sigma$ and the
concentrations.  In Fig.\ref{fig3} we show the lowest sub-band for
50-50\% alloying with different degrees of SRO in two dimensional
strips of width $M=20$. In spite of the finite size, the three cases
already display two dimensional features characteristic of the
infinite limit. The DOS structure changes dramatically as a function
of the local order, with highly localized states near the random and
ordered limits. The upper figure is close to the segregation limit
and resembles the pure case shown in Fig.\ref{fig1}. For this regime,
the electronic wave function may percolate through atoms of the same
type in very large islands.  In contrast with one dimension, the
above properties are typical of the two dimensional topology, with
many different percolation paths for extended electron trajectories.

For three dimensions, the finite section is of two dimensional
character and SRO properties are handled with more parameters that
include the possibility of closed path correlations. This is done
using the Kikuchi method\cite{Kikuchi} and is described in the next
Section.

\section{The inclusion of magnetic fields and SRO properties in
three-dimensional geometries}

Following ideas developed in Finite-Size Scaling methods,
\cite{MacKinonn} our system is taken as a bar in three dimensions,
with finite dimensions in the $x$ and $y$ directions, and infinite
extension along the $z$ direction.  Periodic boundary conditions are
imposed for the finite section of the system. The magnetic field is
applied parallel to the $z$ axis.

We model our system using an Alloy Tight Binding (ATB) Hamiltonian:
\begin{equation}
{\cal H}=\sum_{<i>}E_i\left| i\right\rangle \left\langle i\right|
+\sum_{<ij>}Ve^{i\phi _{ij}}\left|i\right\rangle \left\langle
j\right| +h.c.,
\label{H}
\end{equation}
where off-diagonal elements are taken different from zero only for
nearest neighbors, and $\phi _{ij}$ is a phase that represents the
effect of the magnetic field in an elementary cell with the lattice
parameter as a side (Peierls substitution\cite{peierls}). For
simplicity, we choose the hopping amplitude as real along the $x$
axis ($\phi =0$), and complex along the $y$ axis. The phase
difference between two successive bonds is related to the magnetic
flux enclosed in an elementary cell,
\begin{equation}
\phi _{n+1}-\phi _n=\frac e{\hbar c}\oint \overrightarrow{A}\cdot
\overrightarrow{ds}\quad ,  \label{flux}
\end{equation}
where $\overrightarrow{A}$ is the associated electromagnetic vector
potential. Periodic boundary conditions for the finite section of the
system induce a quantization of the magnetic flux that depends on the
size of the sample. One can get rid of the above constraint by
choosing free ends, but in this case finite size effects decay more
slowly.

As mentioned in the previous Section, we studied the limit
corresponding to the Quantum Site Percolation case, where
$|E_A-E_B|>>V$. SRO is treated using the Kikuchi method with closed
square diagrams as highest order correlations.\cite{Kikuchi} This
approximation is known to correctly describe the two dimensional
topology. For each square we have 16 different configurations. These
configurations are displayed in Fig.\ref{fig4}, with the respective
probabilities ($z_i^{^{\prime }}s$) and degeneracies ($\beta
^{^{\prime }}s$).

Normalization conditions require: 
\begin{equation}
c_A+c_B=1 \   \label{cA}
\end{equation}
for concentrations, 
\begin{eqnarray}
p_{AA}+p_{AB} &=&c_A  \nonumber \\
p_{BB}+p_{AB} &=&c_B \   \label{p}
\end{eqnarray}
for pairs, and 
\begin{eqnarray}
p_{AA} &=&z_1+2z_2+z_3  \nonumber \\
p_{AB} &=&z_2+z_3+z_4+z_5  \label{pz} \\
p_{BB} &=&z_3+2z_5+z_6  \nonumber
\end{eqnarray}
for squares.

In all, we get 5 independent parameters, which are chosen to be
$p_{AB}$, $z_3$, $z_4$, $\xi _1$ and $\xi _2$, with the last two
defined as:
\begin{eqnarray}
\xi _1 &=&c_A-c_B  \nonumber \\
\xi _2 &=&z_2-z_5.  \label{lro}
\end{eqnarray}
While the first three parameters ($p_{AB},z_3,z_4$) are related to
correlations of short range, the last two (\ref{lro}) handle long
range order.

Dependent variables are written as: 
\begin{eqnarray}
c_A &=&(1+\xi _1)/2  \nonumber \\
c_B &=&(1-\xi _1)/2  \nonumber \\
p_{AA} &=&(1+\xi _1-2p_{AB})/2  \nonumber \\
p_{BB} &=&(1-\xi _1-2p_{AB})/2  \nonumber \\
z_1 &=&(1+\xi _1-4p_{AB}+2z_4-2\xi _2)/2 \\
z_2 &=&(p_{AB}-z_3-z_4+\xi {2})/2  \nonumber \\
z_5 &=&(p_{AB}-z_3-z_4-\xi _2)/2  \nonumber \\
z_6 &=&(1-\xi _1-4p_{AB}+2z_4+2\xi _2)/2\   \nonumber
\end{eqnarray}
Within the approximation adopted, a given configuration of the finite
section is obtained by joining squares. The corresponding probability
is obtained as a product of conditional square probabilities, where
the constraint is that two squares are linked by a common bond. This
approach has been devised in analogy to Markovian
chains,\cite{markov} but with the square as the basic correlation.

As it will be seen in the next Section, our method allows to include
local properties under the coupled effects of variable disorder and
application of a magnetic field.

\section{Numerical Results and Discussion}

In the examples that follow, we consider 50-50\% alloying and several
degrees of local order. The site energies are taken to be $E_{A}=0$
and $E_{B}=1000$, in units of the hopping term $V$. In all the
figures, we only display the lowest sub-band, the one centered at
$E_{A}$. Fig.\ref{fig5} shows examples calculated in the absence of a
magnetic field: (a) near the ordered limit; (b) random disorder; and
(c) near the limit of segregation.  The behavior is similar to the
one shown previously for two dimensions. Near the segregated limit
the DOS for each sub-band resembles the one of the pure case,
suggesting the presence of extended electronic states. On the other
hand, when the system is near the ordered limit, the DOS shows the
typical structure of localized states, as it should be expected for
two kinds of atoms with very different site energies. In this
instance, the probability for one state to propagate is nearly zero,
due to almost infinite barriers.  The authors are currently
calculating other cases where the centroids approach to each other,
and the bands overlap. In this situation, quantum diffusion is
possible by means of tunneling through high but finite barriers, and
a delocalization onset is expected as a function of the band
separation.\cite{mello1} To illustrate the role of local order, in
Fig.\ref{fig6} we display the DOS values for three nearby energy
levels as a function of the correlation $\sigma $, in two dimensions.
One of the levels corresponds to a sharp peak in the middle of the
band in the lowest part of Fig.\ref{fig3}, for $\sigma =0$.

Size effects are pictured in Fig.\ref{fig7} for the random alloy. It
is worth noting that the DOS of finite systems already present the
whole structure of the infinite case, even for small transverse
sections. This seems to be characteristic of the bar geometry, where
one of the dimensions is infinite and the bulk limit is extrapolated
with not too big cross sections. Due to numerical limitations, the
next examples in the presence of a magnetic field, are calculated in
finite geometries only (sections of 12 atoms).

We now analyze the role of the magnetic field. In Fig.\ref{fig8} we
show the DOS for a case near the segregation limit. An oscillatory
behavior of the DOS with the field is observed, and the figure
depicts cases up to half the period in the flux. The values are
adjusted to periodic boundary conditions.  The effect is paramount at
the center of the sub-band, with the appearance of a gap for $\Delta
\phi =\pi $. The phenomenon is due to destructive interference of the
electronic paths in the presence of the field. The band width is also
affected, with a striking reduction at half a period.

On the other hand, tiny effects are observed near the random and
ordered limits, with no sensitive changes in the structure of the DOS
as a function of the field. Other cases with different concentrations
were also calculated. As a general rule, the DOS structure is more
affected by the field near the so called {\em weak localization}
regime, when the concentration of one of the species is small.
Localization and delocalization of electronic states may be
interpreted as pure quantum effects due to destructive or
constructive interference of the different electronic paths. This
interference is shifted by the application of a field.  In
Fig.\ref{fig9}, we show the effect of the magnetic field on a
particular level as a function of short-range correlations. The
figure displays a trajectory in our parameter space, ranging from the
local ordered limit to the segregated case. This trajectory, for
short, is labeled with the symbol $\sigma $. The figure pictures in a
different way our comments above: the DOS oscillates as a function of
the field, but the effect predominates in the vicinity of the
segregated limit (for $0.5\leq \sigma \leq 1$ in the figure).

In summary, it has been shown that the method proposed is numerically
efficient, reproduces the DOS in important limits, and predicts new
behaviors as function of the SRO and magnetic field. To theoretically
investigate localization properties, one needs to calculate the
electron wave function, or get indirect information {\em via} the so
called localization length. In two and three dimensions, most of the
works in the literature concentrate around simulations or other
numerical techniques. As a common characteristic, they are based on
Finite Size Scaling methods, but most of the cases investigated
correspond to random disordered systems, represented generally by the
Anderson Model or the Quantum Percolation
Model\cite{Kramer,Soukoulis} Our results, in spite that we calculate
the average {\em DOS}, suggest that {\em SRO effects may play an
important role in the localization of electronic states}. A phase
diagram in which the degree of localization is displayed, may depend
not only on the concentrations, but also on the SRO parameters.  In
order to illustrate this issue, we suggest numerical computations of
the localization length in systems like the ones considered here,
with equal concentrations of the species, very different site
energies, and variable local order properties. Changes in the
localization properties will then be entirely ascribed to short-range
correlations.

The applied magnetic field strongly changes the electronic properties
in the {\em weak localization} limit. On the other hand, in the
opposite regime, the effects of the field on the average DOS are
negligible. Huge magnetic fields are involved in the above phenomena
due to the small size of the elementary cell.  However, the effects
predicted here may be brought to a range susceptible of observation
in the case of superlattice structures, or in geometries where the
electronic paths enclose macroscopic areas that contribute to the
flux. We note that the present state of the art in nanostructure
technology allows the fabrication of modulated 2D systems with
lattice parameters of the order of 100 nm or more (see for instance
Ref.\onlinecite{nano}).

\newpage \acknowledgments

The authors are grateful to Professors D. Gottlieb, B. Laks, N.
Majlis, and P. Schulz for helpful discussions. This work was
partially financed by a grant from {\em Vitae} (Brazil, Project No.
B-11487/2B9221). One of the authors (G.G.C.) also acknowledges
support from {\em Conselho de Desenvolvimento Cient\'{\i}fico e
Tecnol\'ogico} (CNPq, Brazil, Project No.301221/77-4).

\appendix

\section{Exact calculation for the DOS in two dimensions}

The DOS for the pure cubic lattice in $d$ dimensions was given in
Section II by eq.(\ref{DOS}). We show explicitly this result for the
two-dimensional case.

If we consider that states $\{|m,n>\}$ represent the basis in the
site representation, where the indexes are integers along the $x$ and
$y$ axes respectively, the Hamiltonian reads:
\begin{eqnarray}
{\cal H}=\sum_{<m,n>}E_{mn}\left| m,n\right\rangle \left\langle
m,n\right| &+&\sum_{<n>}t_{x}\left| m,n\right\rangle \left\langle
m,n+1\right| +h.c.
\nonumber \\
&+&\sum_{<m>}t_{y}\left| m,n\right\rangle \left\langle m+1,n\right|
+h.c.
\label{H1}
\end{eqnarray}
Taking the Fourier transform in relation to the $x$ direction: 
\begin{equation}
\left| m,n\right\rangle
=\frac{1}{\sqrt{N}}\sum_{<k_{0}>}e^{ikn}\left| m,k\right\rangle ~~,
\end{equation}
we get: 
\begin{eqnarray}
{\cal H} &=&\sum_{m,k}(E_{0}+2t_{x}\cos k)\left| m,k\right\rangle
\left\langle m,k\right| +  \nonumber \\
&&\;\;\sum_{m,k}t_{y}(\left| m,k\right\rangle \left\langle
m+1,k\right| +\left| m+1,k\right\rangle \left\langle m,k\right| )\ .
\end{eqnarray}
The Green's function in the basis $\{|m,k>\}$ is given by:
\cite{Economou}
\begin{equation}
{\bf G}(z)=\sum_{m,k}\frac{\left| m,k\right\rangle \left\langle
m,k\right| }{z-E(k,m)}, 
\end{equation}
where $E(k,m)$ are the eigenvalues: 
\begin{equation}
E(k,m)=E_{0}+2t_{x}\cos k+\lambda _{m}=E_{x}(k)+\lambda _{m},
\end{equation}
with: 
\begin{equation}
E_{x}(k)=E_{0}+2t_{x}\cos k,
\end{equation}
and $\lambda _{m}$ corresponding to the eigenvalue of an {\em
effective} chain with site energy $E_{0}+2t_{x}\cos k$ and hopping
$t_{y}$ between first neighbors along the $y$ direction.

The Green's function can be rewritten as: 
\begin{equation}
{\bf G}(z)=\sum_{m,k}\frac{\left| m,k\right\rangle \left\langle
m,k\right| }{z-E_{x}(k)-\lambda _{m}}\ , 
\end{equation}
with the diagonal element: 
\begin{equation}
\left\langle m,n\right| {\bf G}(z)\left| m,n\right\rangle
=\frac{L}{2\pi N}
\int_{BZ}{dk\frac{1}{z-\lambda _{m}-E_{x}(k)}}
\end{equation}

The DOS and the Green's function satisfy the relation\cite{Economou}: 
\begin{equation}
{\cal D}(E)=-\frac{1}{\pi }Im\sum_{m,n}\left\langle m,n\right| {\bf
G} (E)\left| m,n\right\rangle ,
\end{equation}
with: 
\begin{equation}
\frac{1}{NM}\sum_{m,n}\left\langle m,n\right| {\bf G}(z)\left|
m,n\right\rangle =\frac{1}{M}\sum_{m}\frac{1}{2\pi }\int_{-\pi }^{\pi
}{dk \frac{1}{z-\lambda _{m}-E_{0}-2t_{x}\cos (k)}}\ ,
\end{equation}
and subsequently the DOS is given by: 
\begin{equation}
{\cal D}(E)=\frac{1}{M}\sum_{\lambda _{m}} n(\lambda _{m})~{\cal
D}_{1dim}(E-\lambda _{m}).
\end{equation}
where ${\cal D}_{1dim}(E-\lambda _{m})$ is the DOS of an infinite
linear chain centered at $\lambda _{m}$, and $n(\lambda _{m})$ is the
degeneracy of the eigenvalue.


\begin{figure}[tbp]
\caption{DOS, in arbitrary units, for a pure two-dimensional system:
(a) upper figure, the exact two dimensional infinite sample; (b)
middle, a strip with the finite transverse length $M=500$; (c) lower,
a strip with the finite transverse length $M=100$. In all the cases,
we take $E_A=E_B=0$ and $V_{ij}=1$, with energy steps equal to
$5.10^{-4}$.}
\label{fig1}
\end{figure}

\begin{figure}[tbp]
\caption{DOS, in arbitrary units, for the pure three-dimensional
system: (a) upper figure, the exact three dimensional case obtained
as the convolution of a two-dimensional DOS with a one-dimensional
one; (b) middle, bar with $N \times M$ = 48; (c) lower, a bar with $N
\times M$ = 24. For all the examples, $E_A=E_B=0$ and $V_{ij}=1$,
with energy steps equal to $5.10^{-4}$.}
\label{fig2}
\end{figure}

\begin{figure}[tbp]
\caption{DOS, in arbitrary units, for the square lattice with several
degrees of SRO. All the cases shown in the figure were calculated for
infinite stripes of width $M=20$ chains. Concentrations are equal for
both species and the Cowley parameter is given in each figure. Other
parameters are $E_{A}=0$, $E_{B}=1000$, $V=1$, with energy steps
equal to $5.10^{-2}$.  We only display the lower sub-band: (a) upper
figure, $\sigma=0.8$, tendency to segregation; (b) middle,
$\sigma=0$, random alloy; (c) lower, $\sigma=-0.8 $, tendency to
ordering.}
\label{fig3}
\end{figure}

\begin{figure}[tbp]
\caption{ Possible square configurations for two kinds of atoms
(filled and empty circles). We denote by $z_i$'s the probabilities of
each configuration, and the numbers indicated in the last column are
the respective degeneracies.}
\label{fig4}
\end{figure}

\begin{figure}[tbp]
\caption{DOS, in arbitrary units, for the three-dimensional
disordered case with several degrees of SRO. All the examples
correspond to the infinite system, and were obtained by the
convolution technique explained in the text. The concentrations of
both species are equal, and as in previous examples, we use
$E_{A}=0$, $E_{B}=1000$, $V=1$, with energy steps equal to
$5.10^{-2}$. We only display the lower sub-band for the cases: (a)
tendency to order; (b) random alloy; (c) tendency to segregation.}
\label{fig5}
\end{figure}

\begin{figure}[tbp]
\caption{DOS, in log scale, for the two dimensional case, for three
nearby energy points, as a function of the short-range parameter
$\sigma$.}
\label{fig6}
\end{figure}

\begin{figure}[tbp]
\caption{Size effects for the random disordered binary alloy for the
three dimensional case (case (b) of Fig.\ref{fig5}). For finite
systems we use the following correlation parameters: $p_{AB}=1/4$,
$z_3=z_4=1/16$, and $\xi_1=\xi_2=0$. Note that the lower DOS, for a
bar of cross section of 12 atoms, already presents, except for the
amplitude of some peaks, the main structure of the infinite system
density.}
\label{fig7}
\end{figure}

\begin{figure}[tbp]
\caption{DOS, in arbitrary units, for a three-dimensional bar
($N\times M=12$) for different values of the applied magnetic field.
The case depicted corresponds to a disordered binary alloy near the
segregation limit for equal concentrations of both species.
Correlation parameters are $p_{AB}=0.05 $, $z_3=z_4=0$, and
$\xi_1=\xi_2=0$. We only show the lower sub-band and the magnetic
flux is indicated in each figure. Due to periodic boundary conditions
for the finite section of the bar, the magnetic phase factor is
quantized in units of $\pi/3$.}
\label{fig8}
\end{figure}

\begin{figure}[tbp]
\caption{ DOS, in arbitrary units, at a particular energy point
($E=3.0$), for a three-dimensional bar ($N\times M=12$), for
different values of the applied magnetic field, as a function of the
short-range order. }
\label{fig9}
\end{figure}

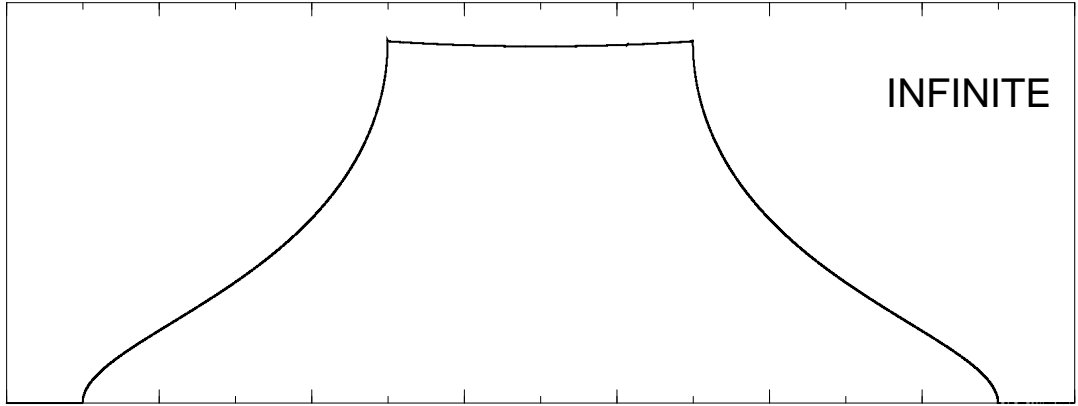
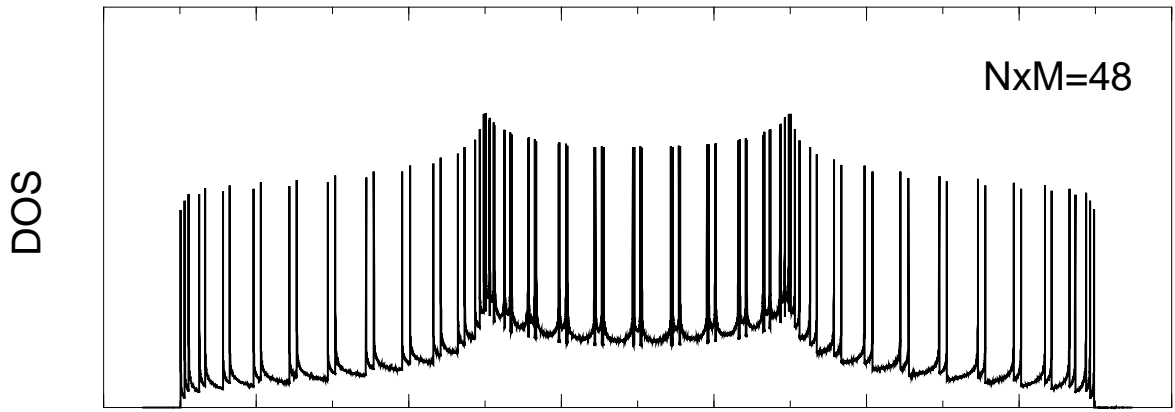
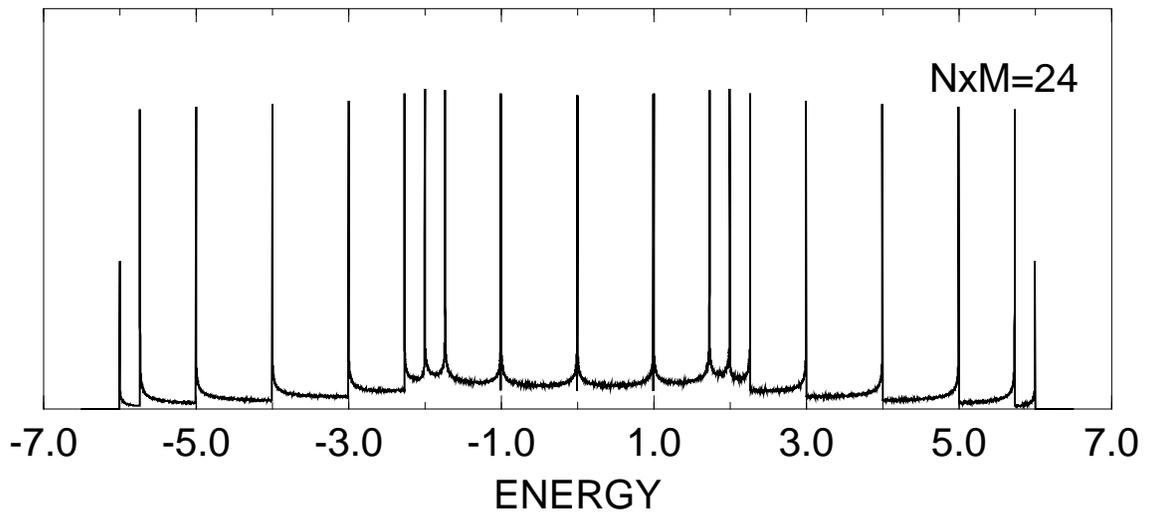
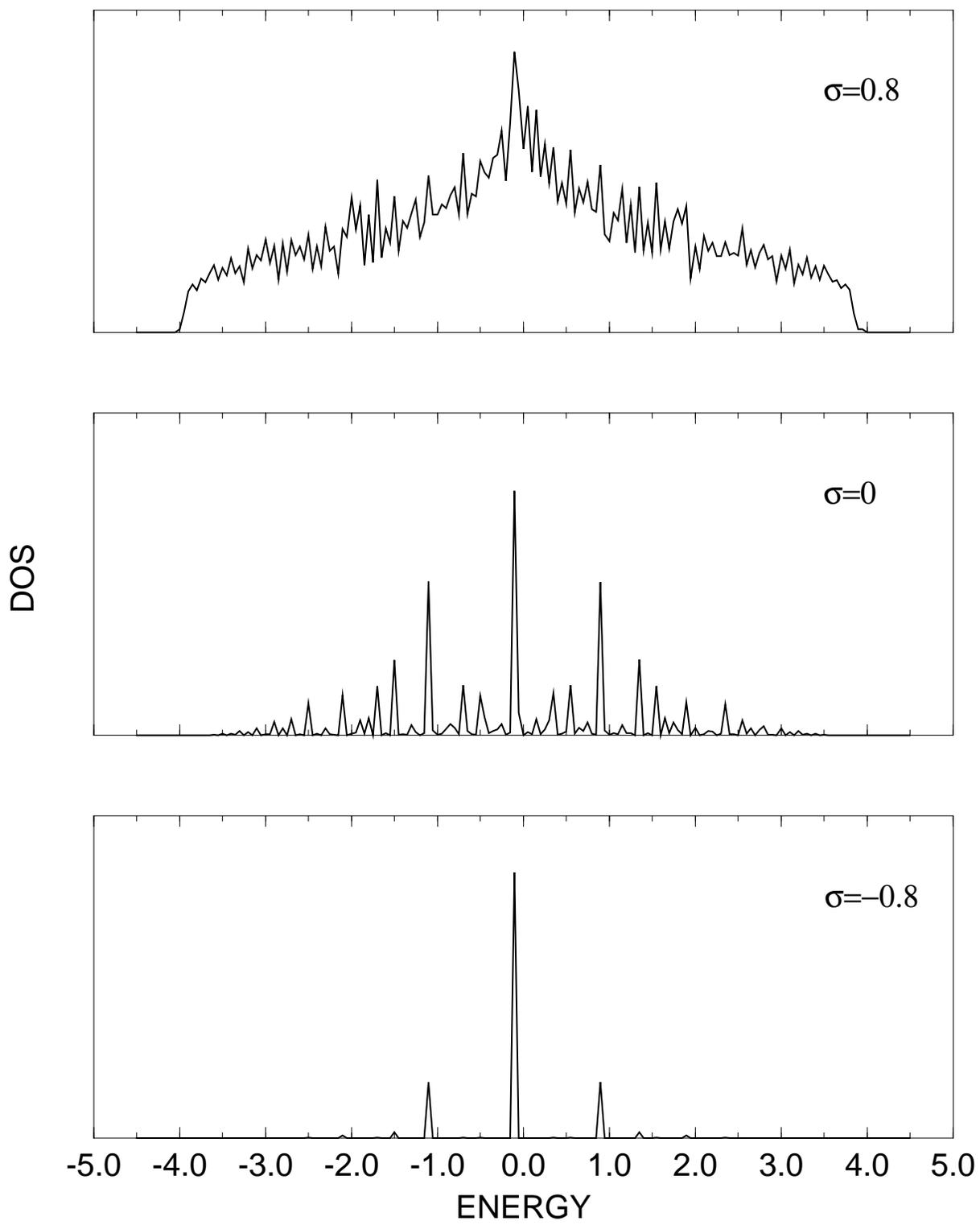
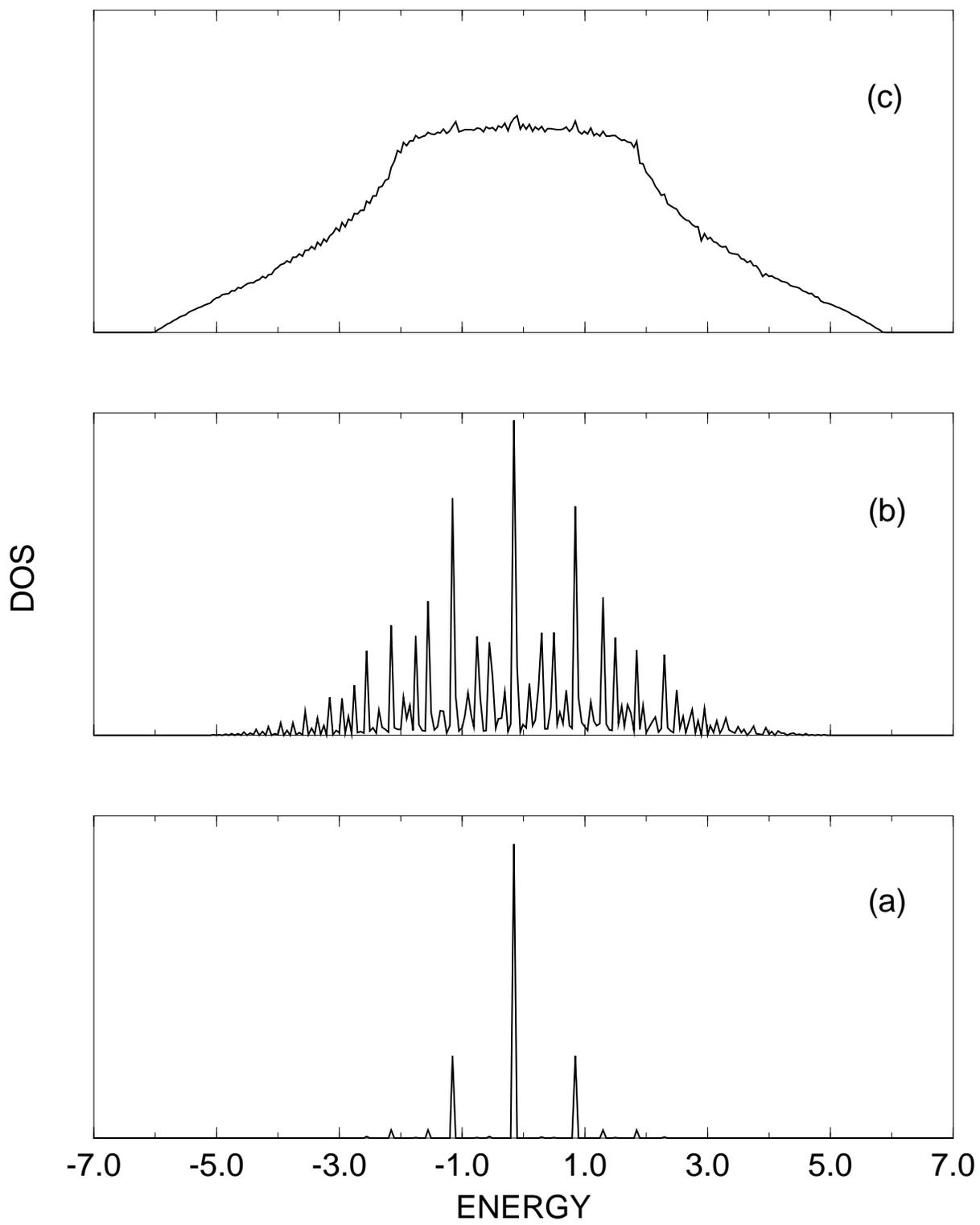
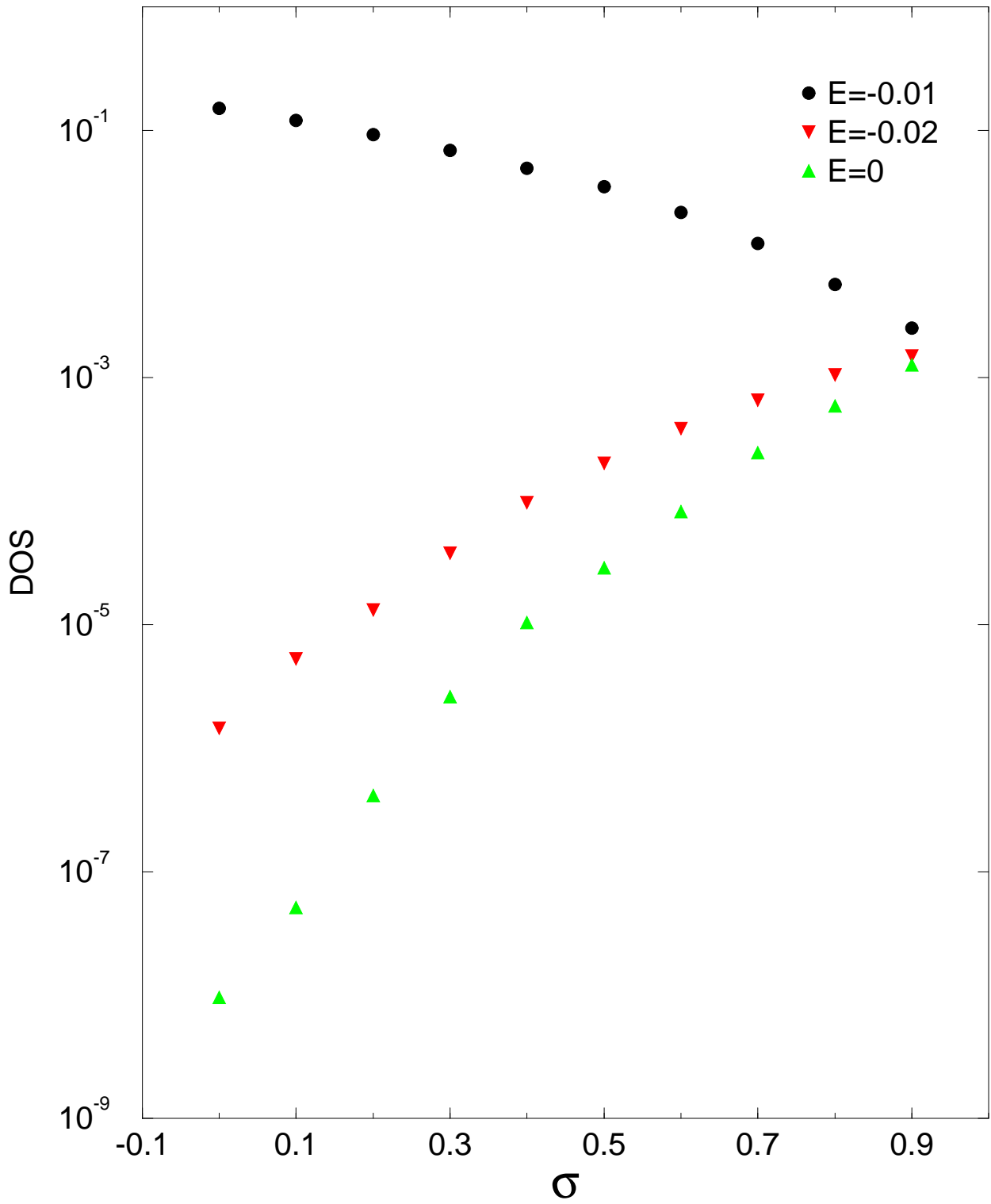
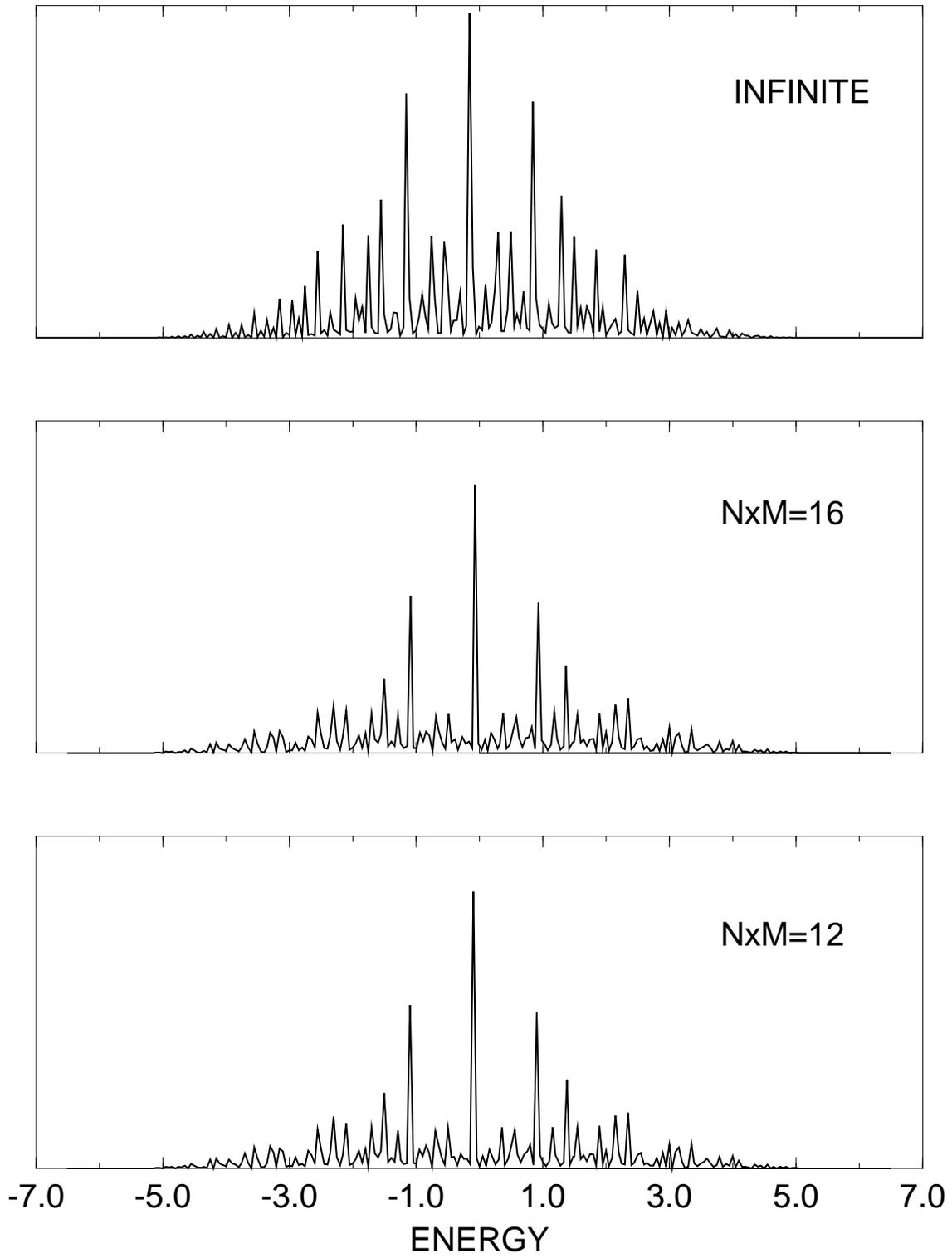
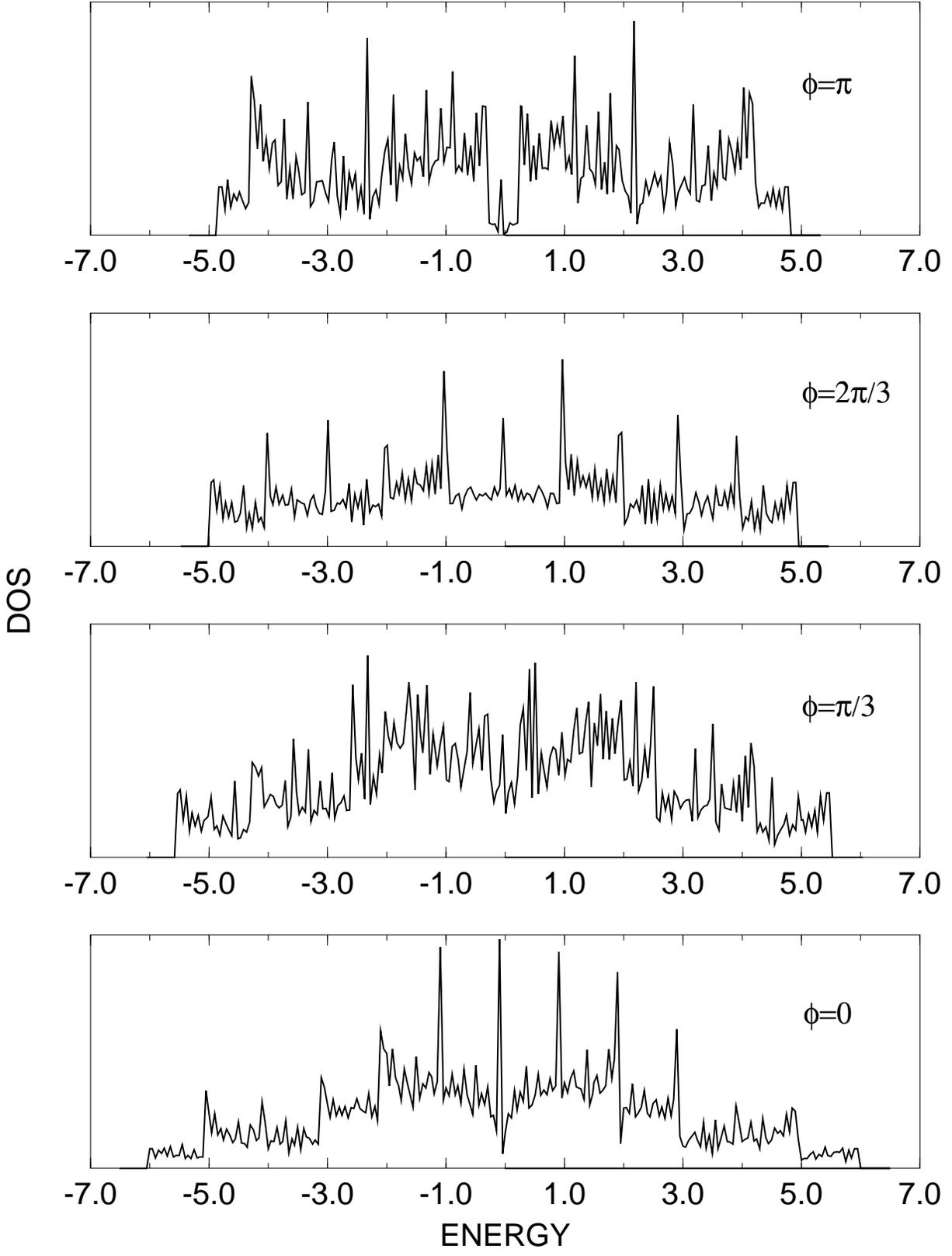
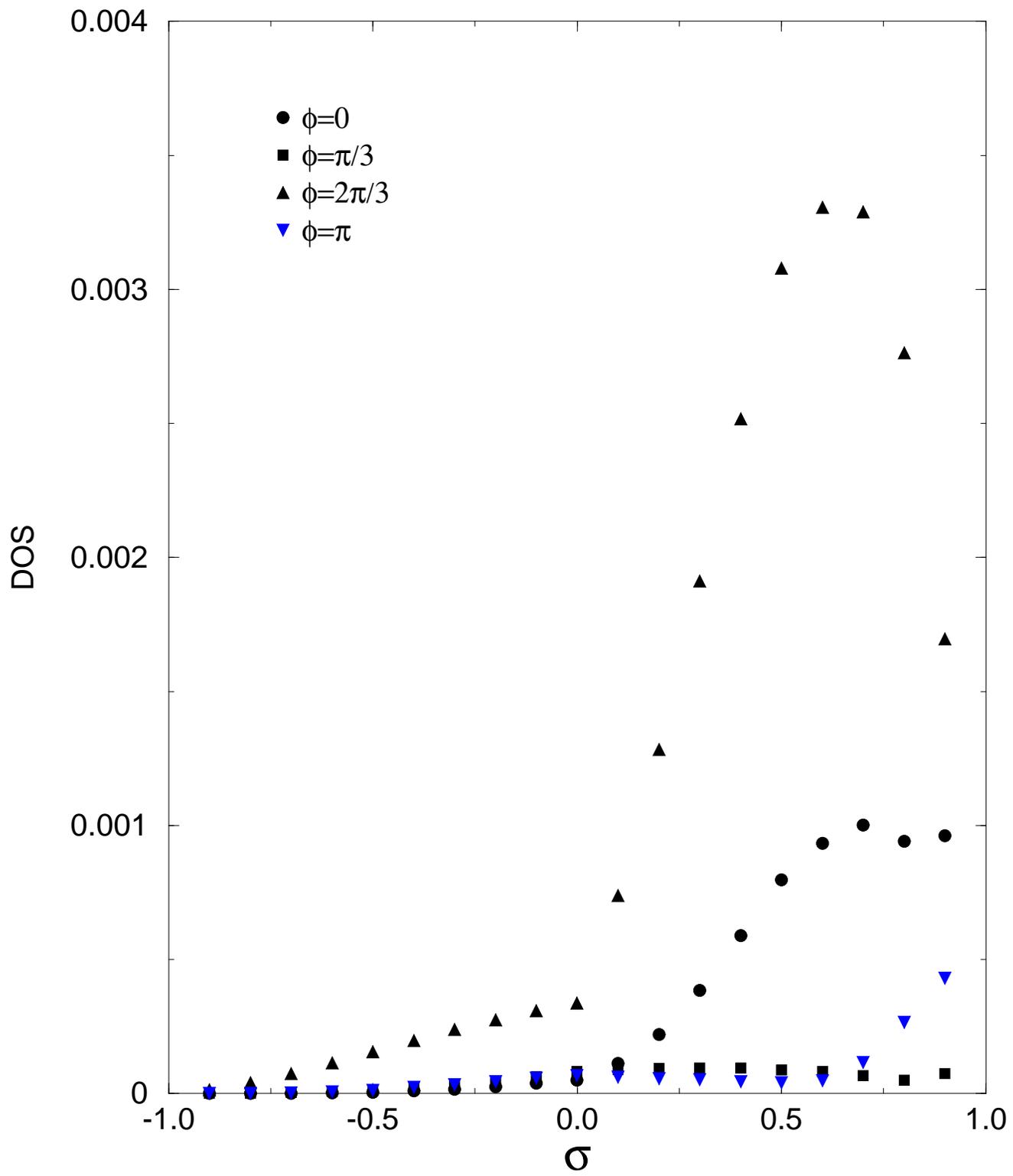
\end{document}